\documentclass[aps,pre,twocolumn,amsmath,amssymb,showpacs]{revtex4-1}
\usepackage{amsmath}
\usepackage{amssymb}
\usepackage{graphicx}
\usepackage{float}
\usepackage{subfigure}
\usepackage{color}
\usepackage{hyperref}
\usepackage{comment}
\usepackage[utf8x]{inputenc}

\begin{document}
\title{Kardar-Parisi-Zhang growth on square domains that enlarge nonlinearly in time}
\author{Ismael S. S. Carrasco}
\email{ismael.carrasco@ufv.br}
\affiliation{Instituto de F\'isica, Universidade de Bras\'ilia, 70919-970, Bras\'ilia, DF, Brazil}
\affiliation{Instituto de F\'isica, Universidade Federal Fluminense, 24210-340 Niter\'oi, RJ, Brazil}
\author{Tiago J. Oliveira}
\email{tiago@ufv.br}
\affiliation{Departamento de F\'isica, Universidade Federal de Vi\c cosa, 36570-900, Vi\c cosa, MG, Brazil}
\date{\today}

\begin{abstract}
Fundamental properties of an interface evolving on a domain of size $L$, such as its height distribution (HD) and two-point covariances, are known to assume universal but different forms depending on whether $L$ is fixed (flat geometry) or expands linearly in time (radial growth). The interesting situation where $L$ varies nonlinearly, however, is far less explored and it has never been tackled for two-dimensional (2D) interfaces. Here, we study discrete KPZ growth models deposited on square lattice substrates, whose (average) lateral size enlarges as $L= L_0 + \omega t^{\gamma}$. Our numerical simulations reveal that the competition between the substrate expansion and the increase of the correlation length parallel to the substrate, $\xi \simeq c t^{1/z}$, gives rise to a number of interesting results. For instance, when $\gamma < 1/z$ the interface becomes fully correlated, but its squared roughness, $W_2$, keeps increasing as $W_2 \sim t^{2\alpha \gamma}$, as previously observed for 1D systems. A careful analysis of this scaling, accounting for an intrinsic width on it, allows us to estimate the roughness exponent of the 2D KPZ class as $\alpha = 0.387(1)$, which is very accurate and robust, once it was obtained averaging the exponents for different models and growth conditions (i.e., for various $\gamma$'s and $\omega$'s). In this correlated regime, the HDs and covariances are consistent with those expected for the steady-state regime of the 2D KPZ class for flat geometry. For $\gamma \approx 1/z$, we find a family of distributions and covariances continuously interpolating between those for the steady-state and the growth regime of radial KPZ interfaces, as the ratio $\omega/c$ augments. When $\gamma>1/z$ the system stays forever in the growth regime and the HDs always converge to the same asymptotic distribution, which is the one for the radial case. The spatial covariances, on the other hand, are $(\gamma,\omega)$-dependent, showing a trend towards the covariance of a random deposition in enlarging substrates as the expansion rate increases. These results considerably generalize our understanding of the height fluctuations in 2D KPZ systems, revealing a scenario very similar to the one previously found in the 1D case.
\end{abstract}

\maketitle

\section{Introduction}
\label{secIntro}

Surface growth is a fascinating research topic, underlying very important technologies (e.g., all those related to thin film deposition), as well as fundamental processes in biological systems and others \cite{barabasi}. In this context, the Kardar-Parisi-Zhang (KPZ) \cite{KPZ} class is of primary importance, being a paradigm of universality of non-equilibrium fluctuations in growth phenomena and a number of other physical systems (such as polymers in random media, driven particles, etc.) \cite{barabasi,healy95,Kriecherbauer2010}. 

If $h(\vec{x},t)$ is a height field defining a translation-invariant surface of lateral size $L$, so these fluctuations can be quantified in terms of the global squared roughness $W_2(L,t) = \left\langle \overline{h^2}(t) - \overline{h}(t)^2 \right\rangle$, where $\overline{\cdot}$ denotes average over the heights of a given surface and $\langle \cdot \rangle$ over different samples (at a given time), respectively. Since the seminal work by Family \& Vicsek (FV) \cite{FV}, it is known that $W_2$ follows a dynamic scaling, with $W_2(L,t) \sim L^{2\alpha} f[\xi(t)/L]$, where $\xi(t) \simeq c t^{1/z}$ is the correlation length parallel to the substrate and the scaling function behaves as $f(x) \sim x^{2 \alpha}$ if $x \ll 1$ and $f(x) = 1$ if $x \sim 1$. Therefore, while $\xi \ll L$, the system is found in a transient growth regime (GR) where the roughness increases asymptotically as $W_2 \sim t^{2\beta}$, with $\beta = \alpha/z$. When the finite system becomes completely correlated (i.e., $\xi \sim L$), $W_2$ stops increasing, but its saturated values scale with the system size as $W_2 \sim L^{2\alpha}$. The one-dimensional (1D) KPZ class is defined by the exponents $\alpha = 1/2$ and $z=3/2$ \cite{KPZ}. From Galilean invariance, one expects also that $\alpha + z = 2$ for KPZ systems in any substrate dimension $d$ \cite{barabasi,KrugAdv}. However, despite 35 years of efforts to calculate these exponents for $d>1$, their exact values are still an open issue. Particularly for the 2D KPZ class, which is our focus here, different analytical approaches to the KPZ equation usually return different exponents \cite{Lassig,Colaiori,Fogedby2,Canet2,Canet3}, and they are not supported by the most accurate numerical estimates available for them \cite{Kelling2011,Pagnani}. Actually, even the outcomes from these large scale simulations of 2D KPZ models do not agree within the error bars. 

In the ideal case of an infinite \textit{flat} substrate ($L \rightarrow \infty$), the system stays in the GR forever, with the height at 1-point of the surface evolving asymptotically according to the ``KPZ ansatz'' $h = v_{\infty} t + s_{\lambda}(\Gamma t)^{\beta} \chi + \cdots$, where the asymptotic growth velocity $v_{\infty}$, the signal of the coefficient $\lambda$ in the KPZ equation \cite{KPZ} $s_{\lambda}$, and the amplitude $\Gamma$ are model-dependent parameters, whereas the growth exponent $\beta$ and the probability density function (pdf) of the fluctuating variable $\chi$ [i.e, the underlying height distribution (HD), $P(\chi)$] are universal. For example, for the 1D KPZ class, $P(\chi)$ is given by the Tracy-Widom (TW) \cite{TW} distribution from a Gaussian orthogonal ensemble (GOE) \cite{Prahofer2000,Calabrese2011}. The 2-pt spatial covariance --- $C_S(r,t) = \langle \tilde{h}(\vec{x}+\vec{r},t)\tilde{h}(\vec{x},t) \rangle$, with $\tilde{h}(\vec{x},t) \equiv h(\vec{x},t)-\overline{h}(t)$ --- is also known for the flat 1D KPZ class and it is related to the Airy$_1$ process \cite{Sasa2005}. No one of these quantities are exactly known for KPZ systems in $d \ge 2$.

For finite substrates (of fixed size $L$), one still obtains the same HDs and covariances above for 1D KPZ systems, provided that $L$ and $t$ are large enough, while $\xi \ll L$, as indeed verified in the celebrated experiments by Takeuchi \textit{et al.} \cite{Takeuchi2011} and in several numerical works \cite{tiago12a,HealyCross,silvia17}. Based on this fact, the asymptotic HDs and covariances have been numerically investigated for the KPZ class in $d=2$ \cite{healy12,tiago13,healy13,Ismael14} and higher dimensions \cite{Alves14,HHTake2015,Alves16}, as well as for the non-linear Villain-Lai-Das Sarma (VLDS) \cite{villain,LDS} class in $d=1$ and $2$ \cite{Ismael16a}. This has been also used to numerically confirm the universality of the HDs for the linear classes by Edwards-Wilkinson (EW) \cite{EW} and Mullins-Herring (MH) \cite{Mullins,*Herring1951} in both $d=1$ and $2$ \cite{Ismael19b}.

Interestingly, if instead of performing the growth on a flat substrate of fixed size, it is started from a seed, such that the surface size $L$ expands linearly in time (i.e., $L \sim t$), the scaling exponents are still the same, but the HDs and covariances change. For instance, for 1D KPZ systems the HDs are given by the TW distribution from a Gaussian unitary ensemble (GUE) in this case, as widely demonstrated analytically \cite{Johansson,Prahofer2000,Sasamoto2010,Amir} and confirmed experimentally \cite{Takeuchi2010,Takeuchi2011} and numerically \cite{Alves11,Alves13,HealyCross,silvia17,Roy}; and the spatial covariance is now related to the Airy$_2$ process \cite{Prahofer2002}. This dependence with the initial condition has also been analytically demonstrated for the EW and MH classes \cite{Ismael19b}, and numerically verified for the 2D KPZ class \cite{healy12,tiago13,healy13,Ismael14}, as well as for the VLDS class \cite{Ismael16a}. Hence, the splitting of universality classes for surface growth into subclasses depending on whether $L$ is fixed or $L \sim t$ is a quite general feature in growth phenomena.

We remark that systems expanding linearly in time never become completely correlated, because $L \sim t$ increases faster than $\xi \sim t^{1/z}$, once $z>1$ \cite{barabasi,KrugAdv}. This leads us to inquire: what happens in the more general case where $L(t)$ varies nonlinearly? A first step to answer this interesting question was recently given by us for the 1D KPZ class, considering the situation where $\left< L \right> = L_0 + \omega t^{\gamma}$ \cite{Ismael19} and by varying $\omega$ and $\gamma$ a very rich scenario for the HDs and spatial covariances was numerically found \cite{Ismael19}. In the present work, we generalize this for 2D KPZ systems, by performing extensive simulations of discrete KPZ models on square lattice substrates whose lateral sizes enlarge as $\left< L_x \right> = \left< L_y \right> = L_0 + \omega t^{\gamma}$. Once again, very interesting behaviors are obtained depending on the expansion rate. For example, the HDs are always given by the 2D counterpart of the TW-GUE distribution for $\gamma > 1/z$, but the spatial covariances are $(\gamma,\omega)$-dependent in this regime. For $\gamma<1/z$ the system becomes fully correlated and the HDs and covariances correspond to those of the steady-state regime of flat interfaces. The roughness, however, keeps increasing as $W_2 \sim t^{2\alpha \gamma}$, allowing us to obtain an accurate estimate for the roughness exponent $\alpha$. For $\gamma \approx 1/z$, we find a family of distributions continuously varying between those for $\gamma < 1/z$ and $\gamma>1/z$ as the ratio $L/\xi$ increases.

\begin{figure}[!t]
	\centering
	\includegraphics[width=8.5cm]{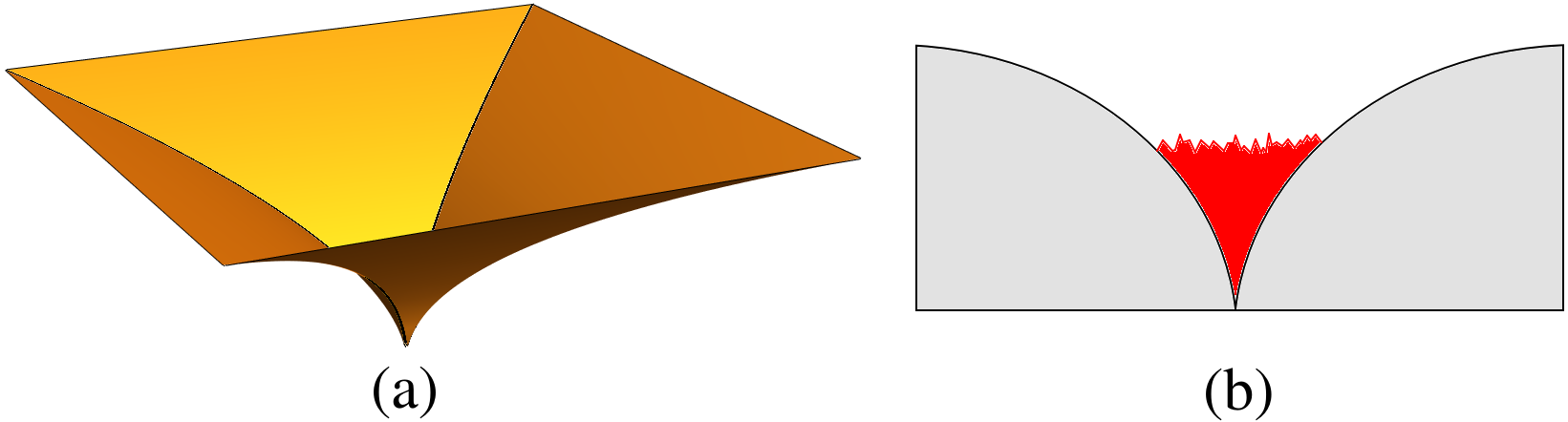} 	
	\caption{(a) Illustration of a hole-like substrate for $\gamma > 1$. (b) Vertical cross-section of a deposit/surface (in red) growing on the substrate shown in (a).}
	\label{fig0}
\end{figure}

It is worth noticing that in the 1D case these nonlinearly expanding systems may represent circular interfaces evolving out of the plane, on the surface of a background space given by a solid of revolution \cite{Ismael19}, but this seems to have no physically realizable analog in 2D. However, if deposition is performed on a hole-like substrate --- with curved and symmetric walls yielding a square horizontal cross-section (see Fig. \ref{fig0}) --- for special aggregation rates (at the deposit, its contour and naked substrate) and wall shapes, it seems possible to obtain square surfaces expanding as in our system as they grow.

The rest of this paper is organized as follows. In Sec. \ref{secMod} we briefly introduce the investigated models and the kinetic Monte Carlo method used to simulate them on expanding substrates. Results for the roughness scaling, HDs, and spatial covariances are respectively presented in Secs. \ref{secResW}, \ref{secResHDs} and \ref{secResCov}. Our final discussions and conclusions are summarized in Sec. \ref{secConc}.

\section{Models}
\label{secMod}

We investigate the restricted solid-on-solid (RSOS) model by Kim \& Kosterlitz \cite{kk} and the single-step (SS) model by Meakin \textit{et al.} \cite{ss1}, which are two workhorses for numerical studies of the KPZ class. In both models, particles are sequentially released toward a horizontal substrate, with $L_x \times L_y$ sites and unitary lattice spacing, at randomly chosen sites. Periodic boundary conditions are used in both directions of the substrate. In the RSOS model, these particles are monomers (with size $1\times 1 \times 1$), and the aggregation at a given site $i$ is accepted only if it does not generate steps larger than 1 at the surface. Namely, the aggregation only occurs if it yields $(h_i - h_j) \leq 1$ for all nearest neighbor (NN) $j$ of site $i$; otherwise, the particle is rejected. In the SS model the particles are vertical dimers, and they only aggregate at a given site, say $i$, if $h_i$ is a local minimum (i.e., if $h_i<h_j$ $\forall$ NN's $j$). 

To study these models on flat substrates, the growth is started with $h_{i}(t=0)=0$ for $i=1,\ldots,(L_xL_y)$ in the RSOS case, while for the SS model a checkerboard initial condition (IC), with $h_{i}(t=0)$ alternating between 0 and 1, is used. In this fashion, the SS surfaces evolve with all local steps satisfying $|h_i - h_j| = 1$. As usual, the time is defined such that we attempt to deposit one monolayer of particles per time unity. This means that, in systems with fixed size, $t \rightarrow t + \Delta t$, with $\Delta t = 1/(L_x L_y)$ after each deposition \textit{attempt}. 

However, here, we are interested in investigating the case where the substrate enlarges isotropically and nonlinearly in time, with $\langle L_x \rangle = \langle L_y \rangle = L_0 + \omega t^{\gamma}$. In order to do this, we follow the method introduced in Ref. \cite{Ismael14} and recently generalized in \cite{Ismael19} to the present case, consisting in stochastically mixing particle deposition with random duplications of lattice rows and columns to make its (average) size variate at a given rate. So, we will consider this rate as $\delta = d \langle L_x \rangle/ d t =  d \langle L_y \rangle/ d t = \gamma \omega t^{\gamma-1}$ and, at each step in our kinetic Monte Carlo (kMC) simulations, we randomly choose one of three events: a particle deposition [with probability $P_{dep} = L_x L_y/(L_x L_y + 2 \delta)$], a row duplication [with probability $P_{dup} = \delta/(L_x L_y + 2 \delta)$] or a column duplication [also with probability $P_{dup}$]. Hence, now $\Delta t = 1/(L_x L_y + 2 \delta)$. The row (column) duplication is performed by randomly sorting a given row $r$ (column $c$) and creating a new and identical one at position $r+1$ ($c+1$), after shifting all rows (columns) at its right-hand-side one position to the right \cite{Ismael14,Ismael19}. In the SS model, we have to duplicate a pair of NN rows or columns to avoid a breakdown of the SS condition $|h_i - h_j| = 1$. So, in this case we use $\delta/2$ in the equations above. 

Although we will present some results in what follows for the growth starting on substrates with size $L_0>0$, to avoid undesired crossovers introduced by this initial size \cite{Ismael14,Ismael18}, most of our simulations will be performed for $L_0=0$. In this case, the system starts with a single site in the RSOS model at the initial time $t_0=(1/\omega)^{1/\gamma}$. Since $L_x$ and $L_y$ have to be even in the SS model, we start its growth with a $2 \times 2$ checkerboard lattice at time $t_0=(2/\omega)^{1/\gamma}$ to mimic $L_0=0$.

\section{Results}

\subsection{Roughness scaling}
\label{secResW}

\begin{figure}[!b]
	\centering
	\includegraphics[height=4.8cm]{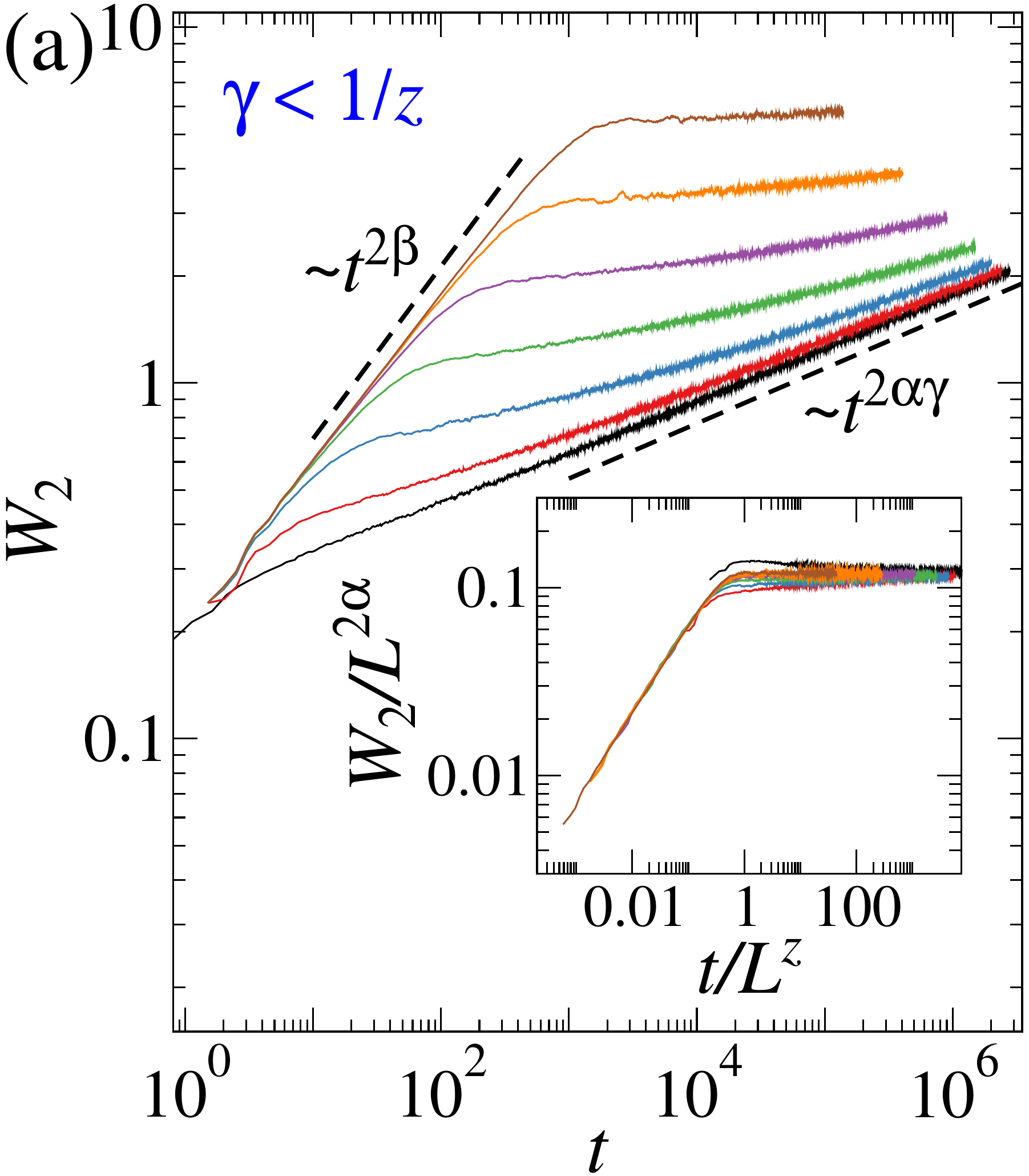} 	
	\includegraphics[height=4.8cm]{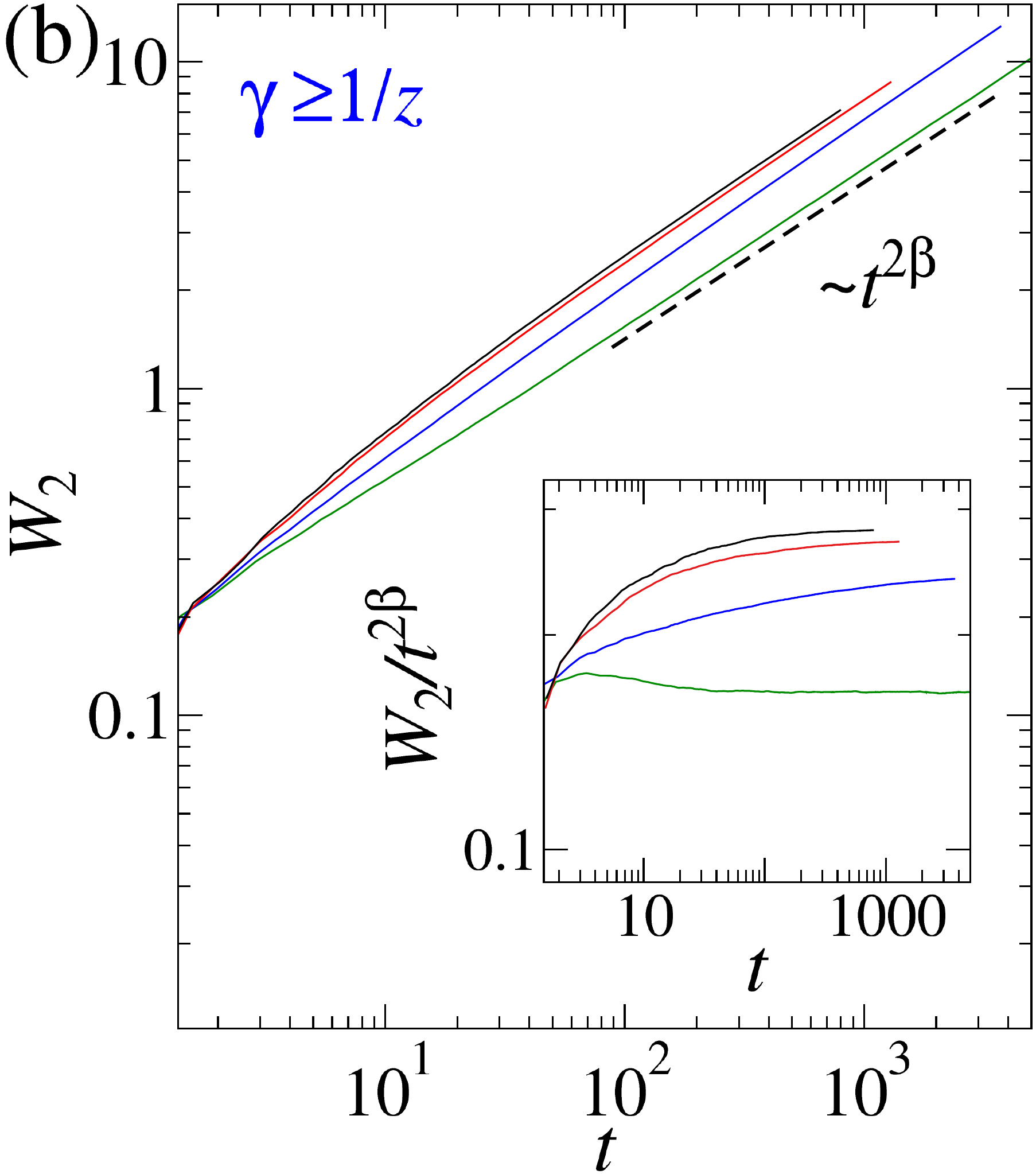}
	\caption{Squared roughness, $W_2$, as a function of time, $t$, for the RSOS model with (a) $\omega=2$, $\gamma=0.2$ and initial sizes (from bottom to top) $L_0=0$, $4$, $8$, $16$, $32$, $64$, and $128$; and (b) $\omega = 2$, $L_0 = 0$ and (from bottom to top) $\gamma = 1/z$, $0.8$, $1.0$ and $1.1$. The insertion in (a) shows the data in the main plot rescaled according to the Family-Vicsek scaling. In (b), the inset displays the temporal variation of the rescaled roughness $W_2/t^{2 \beta}$, where $\alpha = 0.387$ and $\beta = 0.24$ were used.}
	\label{fig1}
\end{figure}

Previous studies on the KPZ \cite{Ismael19,Pastor} and other universality classes \cite{Escudero2009} on 1D substrates enlarging as $L \sim t^{\gamma}$ have demonstrated that, if $L$ enlarges faster than the correlation length, $\xi \sim t^{1/z}$, the roughness increases asymptotically as $W_2 \sim t^{2 \beta}$. Namely, for any $\gamma > 1/z$, the system stays forever in the growth regime (GR), as it does in the widely investigated case of circular interfaces evolving on the plane ($\gamma = 1$). When $\gamma < 1/z$, on the other hand, the system may be found in the GR (with $W_2 \sim t^{2 \beta}$) at short times, but it becomes completely correlated at long times and then $W_2 \sim L^{2\alpha}$. Therefore, the Family-Vicsek \cite{FV} scaling holds also for these 1D expanding systems, but, instead of the saturation observed in fixed-size substrates, the roughness keeps increasing as $W_2 \sim t^{2 \alpha \gamma}$ in the correlated regime. Note that $\gamma = 1/z$ is a kind of ``critical'' situation where the growth and correlated regimes have the same scaling: $W_2 \sim t^{2 \alpha/z} \sim t^{2 \beta}$. 

\begin{figure}[!t]
	\centering
	\includegraphics[height=3.9cm]{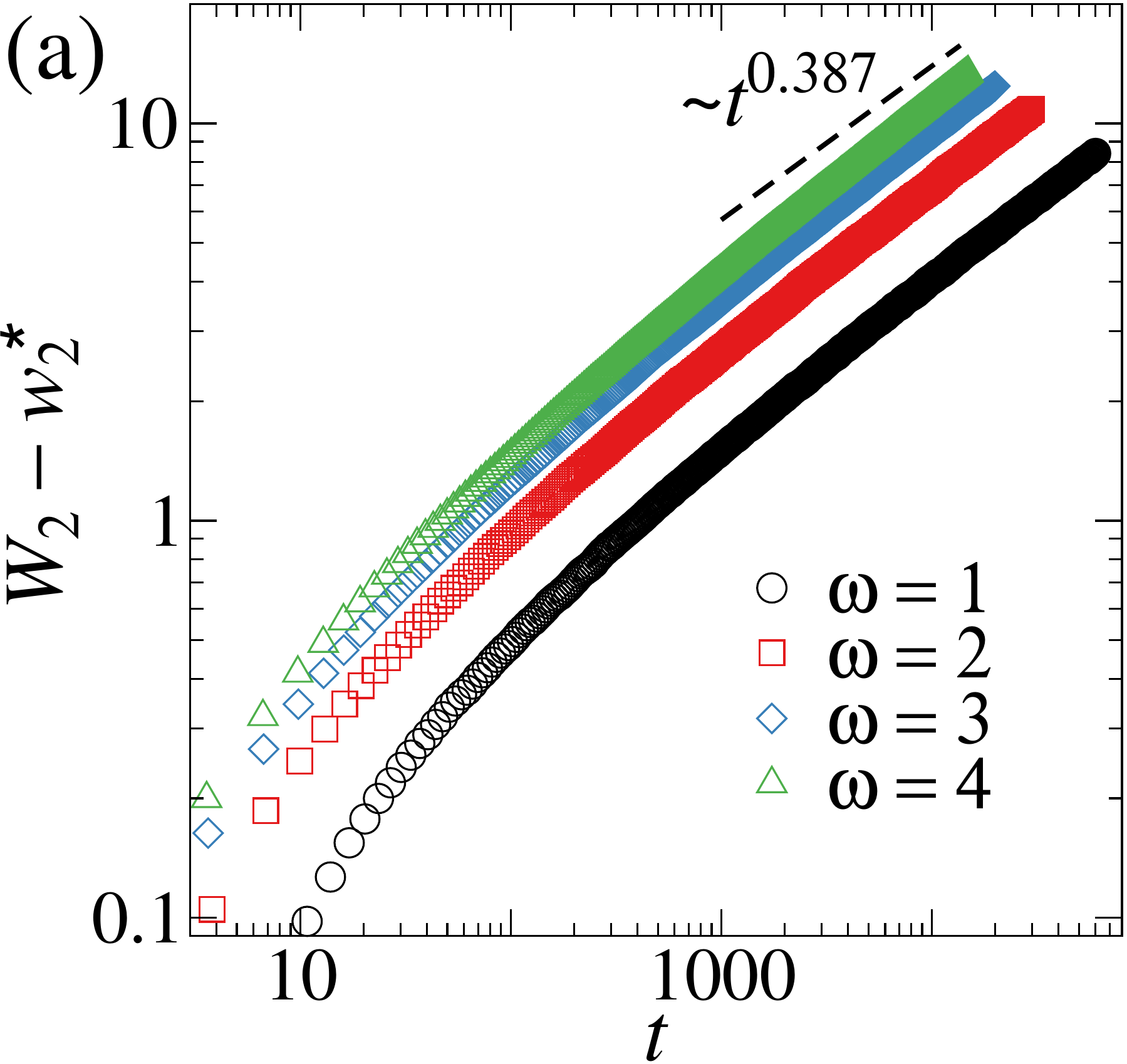} 
	\includegraphics[height=3.9cm]{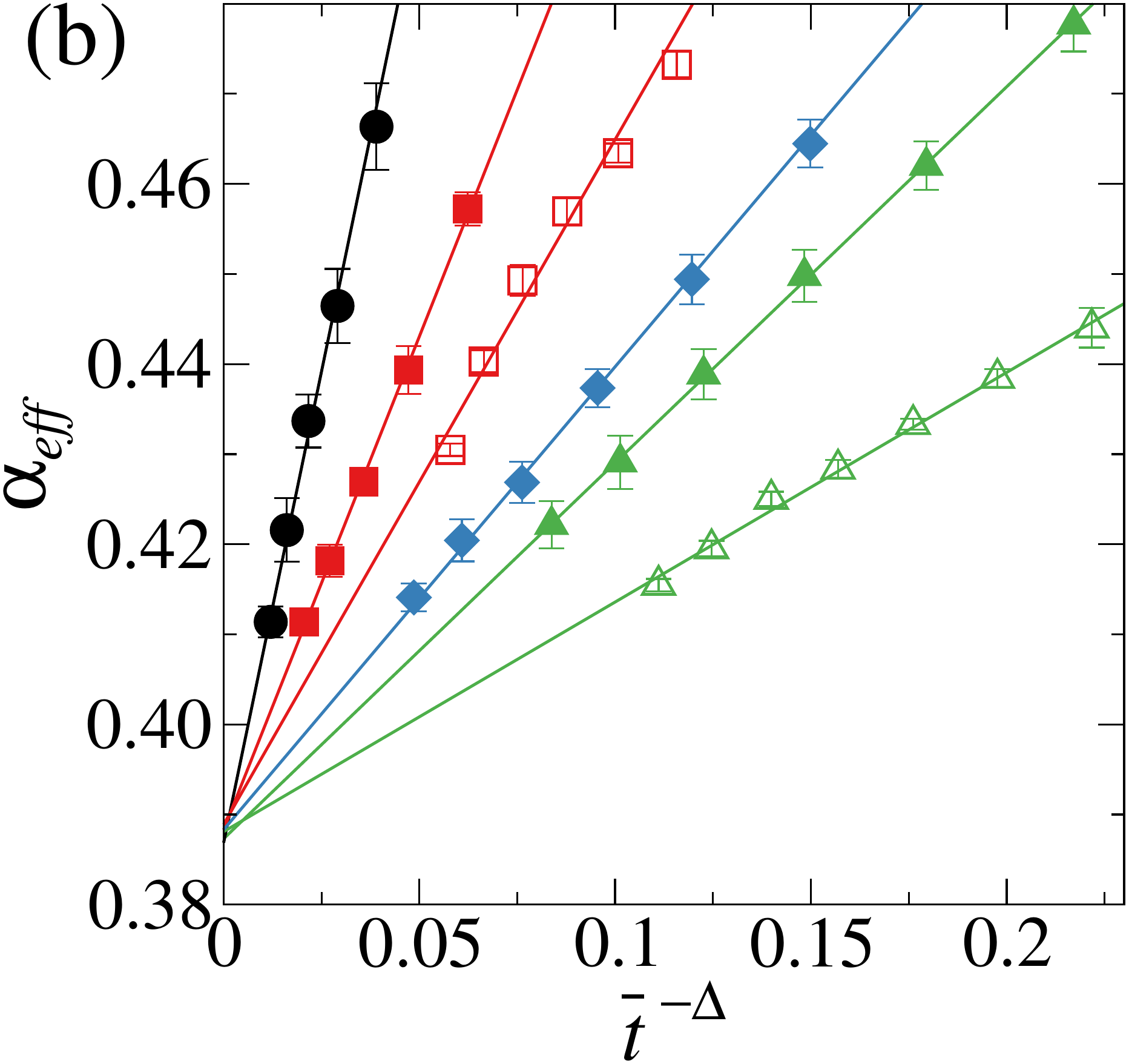}	
	\caption{(a) Squared roughness subtracted of the intrinsic width ($W_2-w_2^*$) as a function of time $t$, for the RSOS model with $\gamma=0.5$ and several values of $\omega$, as indicated by the legend. The dashed line has the indicated slope. (b) Effective roughness exponents $\alpha_{eff}$ against $\bar{t}^{-\Delta}$, for the SS (two upper curves) and RSOS (four lower curves) models, with several values of $\omega$ [following the same symbol and color schema as in (a)] for $\gamma=0.2$ (open) and $0.5$ (closed symbols). The solid lines are linear fits used in the extrapolations. All results here are for an initial size $L_0=0$.}
	\label{fig2}
\end{figure}

As demonstrated in Figs. \ref{fig1} and \ref{fig2} the scenario above is also found for 2D KPZ interfaces deposited on square substrates, whose lateral size $L = \left< L_x \right>=\left< L_y \right>$ varies as $L = L_0 + \omega t^{\gamma}$. In fact, Fig. \ref{fig1}(a) shows an example of the temporal evolution of $W_2$ for $\gamma < 1/z$ --- we will consider here that $z \approx 1.613$ \cite{Pagnani}, so that $1/z \approx 0.620$ --- and two clear scaling regimes are found: $W_2 \sim t^{2\beta}$ at short times (during the GR) and $W_2 \sim t^{2\alpha \gamma}$ asymptotically (in the correlated regime). As expected, by increasing the initial size $L_0$ the duration of the GR augments. Moreover, the crossover to a clean scaling $W_2 \sim t^{2\alpha \gamma}$ becomes very slow for large $L_0$. Despite this, rescaled curves of the roughness $W_2/L^{2\alpha}$ versus $t/L^z$ present a good collapse, as shown in the inset of Fig. \ref{fig1}(a), confirming that the Family-Vicsek scaling is followed by these expanding 2D KPZ surfaces.

Figure \ref{fig1}(b) presents examples of the roughness behavior for $\gamma \geqslant 1/z$ and no crossover to the correlated regime is observed there. Instead, the roughness simply increases approximately as $W_2 \simeq C t^{2\beta}$ at long times, as expected for the GR. It is worth recalling that the substrate expansion yields a correction of type $\theta t^{1-\gamma}$ in the ``KPZ ansatz'' for the 1-point height in the GR, such that $h \simeq v_{\infty} t + s_{\lambda}(\Gamma t)^{\beta} \chi + \eta + \theta t^{1-\gamma}$, where $\eta$ and $\theta$ are expected to be stochastic variables \cite{Ismael19}. While this last term is irrelevant for $\gamma>1$, for $1/z < \gamma \leq 1$ it may introduce important corrections to the roughness scaling. This is consistent with the results in Fig. \ref{fig1}(b), where one indeed observes stronger deviations from the expected scaling for this range of $\gamma$'s. Similarly to what happens in the 1D case, we find evidence here that the scaling amplitude $C=W_2/t^{2\beta}$ presents a small variation with $\gamma$, when $L$ enlarges faster than $\xi$, as indicated by the insertion in Fig. \ref{fig1}(b). A similar dependence is observed also on $\omega$. In the correlated regime found for $\gamma<1/z$, on the other hand, the scaling amplitude $B = W_2/L^{2\alpha}$ seems to be independent of the parameters $\gamma$, $\omega$ and $L_0$. 

Note that the asymptotic behavior of the roughness for $\gamma<1/z$ provides a route to estimate the roughness exponent $\alpha$ from a temporal scaling. In order to do this, it is important to avoid the slow crossover introduced by an initial size $L_0>0$, as seen in Fig. \ref{fig1}(a). So, hereafter we will work only with $L_0 = 0$. Furthermore, as discussed in Ref. \cite{Alves14BD}, an additive and constant correction to scaling, i.e., an intrinsic width, $w_2^*$, is always expected in growing systems; and it is given by the variance, $\langle(\delta h)^2\rangle_c$, of the probability distribution, $P_{i}(\delta h)$, for the height increment $\delta h=h(\vec{x},t+\Delta t)-h(\vec{x},t)$ in each deposition attempt. Indeed, at relatively short times one observes that $\langle(\delta h)^2\rangle_c \rightarrow const. = w_2^*$ \cite{Alves14BD}. For the RSOS model, one has asymptotically that $P_i(0)=1-v_{\infty}$, $P_i(1)=v_{\infty}$ and $P_i(k)=0$ for $k \ge 2$, leading to $w_2^* = v_{\infty} (1 - v_{\infty})$ \cite{Alves14BD}. For the SS model, one has $P_i(0)=1-v_{\infty}/2$, $P_i(1)=0$, $P_i(2)=v_{\infty}/2$ and $P_i(k)=0$ for $k \ge 3$, so that $w_2^* = v_{\infty} (2 - v_{\infty})$. From Ref. \cite{tiago13} one knows that $v_{\infty} = 0.31270$ for the RSOS model and $v_{\infty} = 0.341437$ in the SS case on 2D substrates, which are not affected by the substrate expansion. Thereby, this yields $w_2^* = 0.215$ and $w_2^* = 0.566$ for the RSOS and SS model, respectively.

We remark that previous works analyzing the roughness scaling of these models have never considered the intrinsic width. Since $w_2^*$ is small for them, it has indeed only a mild effect on the scaling if one considers large substrate sizes and long times, such that $W_2 \gg w_2^*$. In our expanding systems, however, the maximum value attained by the squared roughness is $W_2 \sim 10$ and, thus, it is important to account for the intrinsic width. In this way, in the correlated regime we have $W_2 \simeq B t^{2\alpha\gamma} + w_2^*$, so that one has to focus on the scaling of $W_2 - w_2^*$ to estimate the exponent $\alpha$. Examples of the temporal variation of $W_2 - w_2^*$ for the RSOS model are displayed in Fig. \ref{fig2}(a), where one sees that the curves (in log-log scale) are still not so linear, indicating the existence of further corrections to scaling. Similar results are found for other parameters, as well as for the SS model. Therefore, we calculate effective exponents, $\alpha_{eff}$, through the successive slopes of curves of $\log(W_2-w_2^*) \times \log t$. Such slopes were determined in time windows of one decade --- i.e., extending from $t_{min}$ to $t_{max}$, with $t_{max}/t_{min}=10$ --- and to each window we associate a characteristic time $\bar{t}=(t_{min}+t_{max})/2$. By starting with the window for $t_{max}$ equal to the maximum deposition time, we choose the new windows by decreasing both $t_{min}$ and $t_{max}$ by a factor three ($t_{min,max} \rightarrow t_{min,max}/3$). This process is repeated until we get a reasonable number of points to extrapolate.

Examples of the resulting effective exponents are depicted in Fig. \ref{fig2}(b) as a function $\bar{t}^{-\Delta}$, with $\Delta$ being the exponent that best linearizes the data in each case. From linear fits of these data we obtain the extrapolated values (for $t \rightarrow \infty$) summarized in Tab. \ref{TabExp}. It is quite remarkable that these asymptotic estimates, obtained for different growth conditions and models, are so close. In fact, they yield $\alpha = 0.387(1)$, which is in striking agreement with the value estimated by Pagnani \& Parisi \cite{Pagnani} [$\alpha = 0.3869(4)$] using large scale simulations of the RSOS model with multisurface coding. It is worth noticing here that the robustness of the exponents in Tab. \ref{TabExp} is lost if the intrinsic width is disregarded in the scaling, with extrapolated exponents ranging from $0.382$ to $0.399$ being found considering the ``pure'' Family-Vicsek scaling. Anyhow, the average of such exponents gives $\alpha=0.389(5)$, which is not so different from the more reliable result above.

\begin{table}
\caption{Summary of the roughness exponents $\alpha$ obtained from extrapolations of effective exponents as done in Fig. \ref{fig2}(b).}
\begin{tabular}{c c c c c c c}
\hline\hline
$\omega$    & & RSOS ($\gamma = 0.2$)    & &  RSOS ($\gamma = 0.5$) & &  SS ($\gamma=0.5$)  \\
\hline
$1$         & &  0.389                   & &  0.386                 & &  0.387              \\
$2$         & &  0.389                   & &  0.387                 & &  0.388              \\
$3$         & &  0.386                   & &  0.388                 & &  0.386              \\
$4$         & &  0.388                   & &  0.387                 & &  0.385              \\
\hline\hline
\end{tabular}
\label{TabExp}
\end{table}

\subsection{Height distributions}
\label{secResHDs}

Now, we investigate the effect of the nonlinear expansion of the substrate on the height distributions (HDs). All results presented here, and in the next subsection, are for $L_0=0$, so that $L=\langle L_x \rangle = \langle L_y \rangle = \omega t^{\gamma}$. To quantitatively characterize the HDs, we will analyze the adimensional ratios of their first central moments $W_n = \left\langle \overline{(h - \overline{h})^n} \right\rangle$, focusing on the skewness $S = W_3/W_2^{3/2}$ and (excess) kurtosis $K = W_4/W_2^2-3$.

\begin{figure}[!t]
	\centering
	\includegraphics[height=3.9cm]{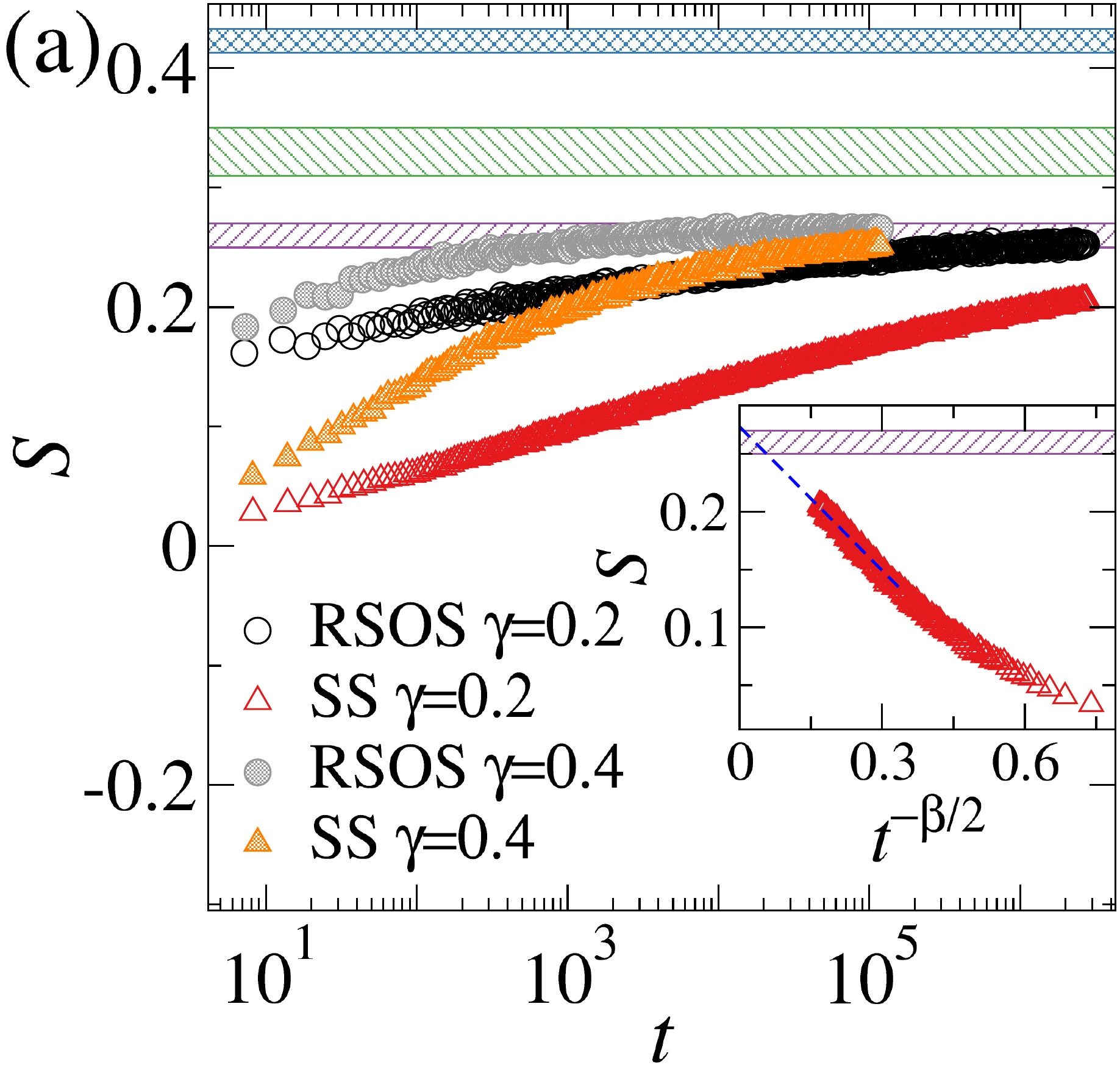} 
	\includegraphics[height=3.9cm]{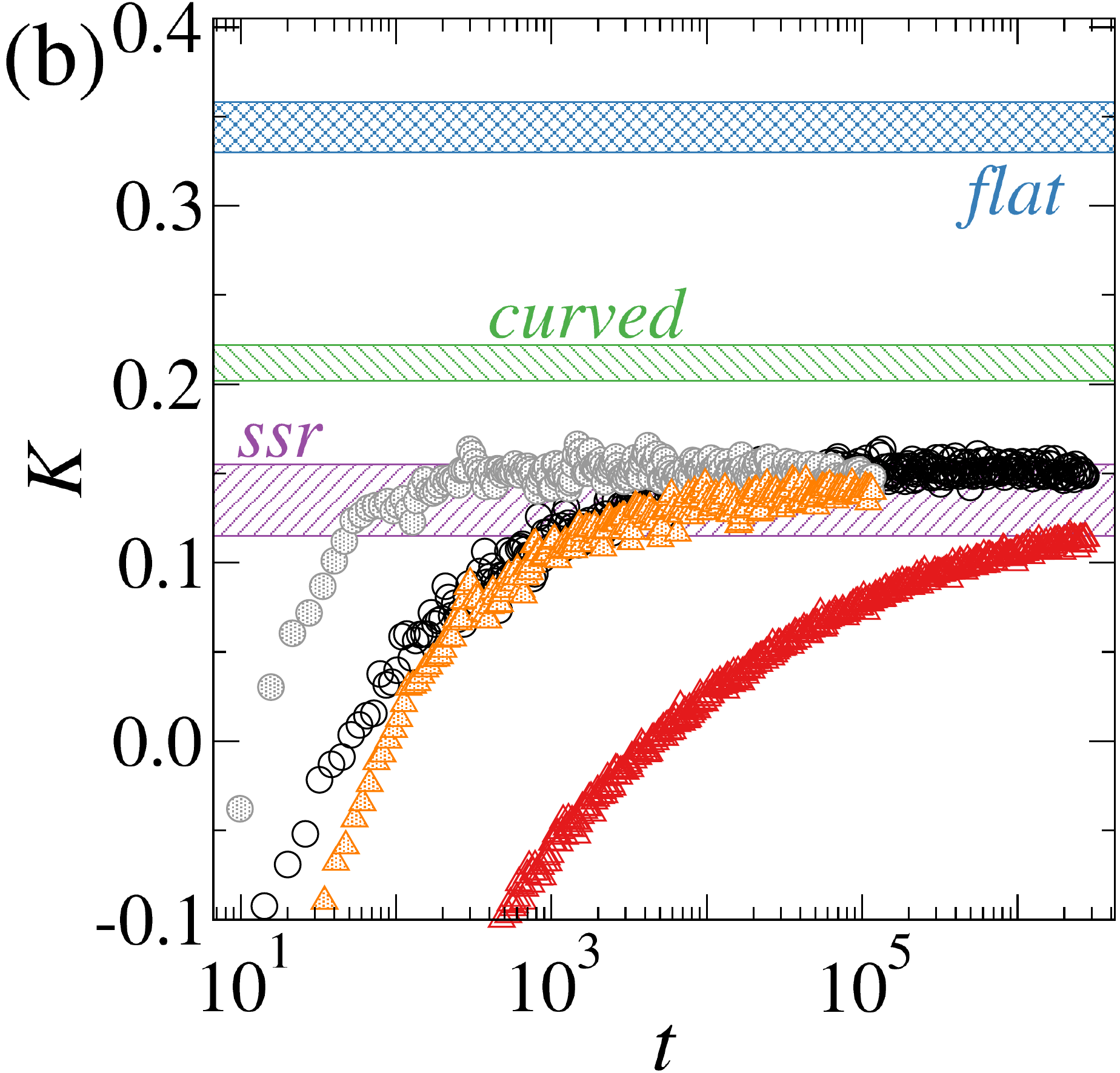}\\
	\includegraphics[height=3.9cm]{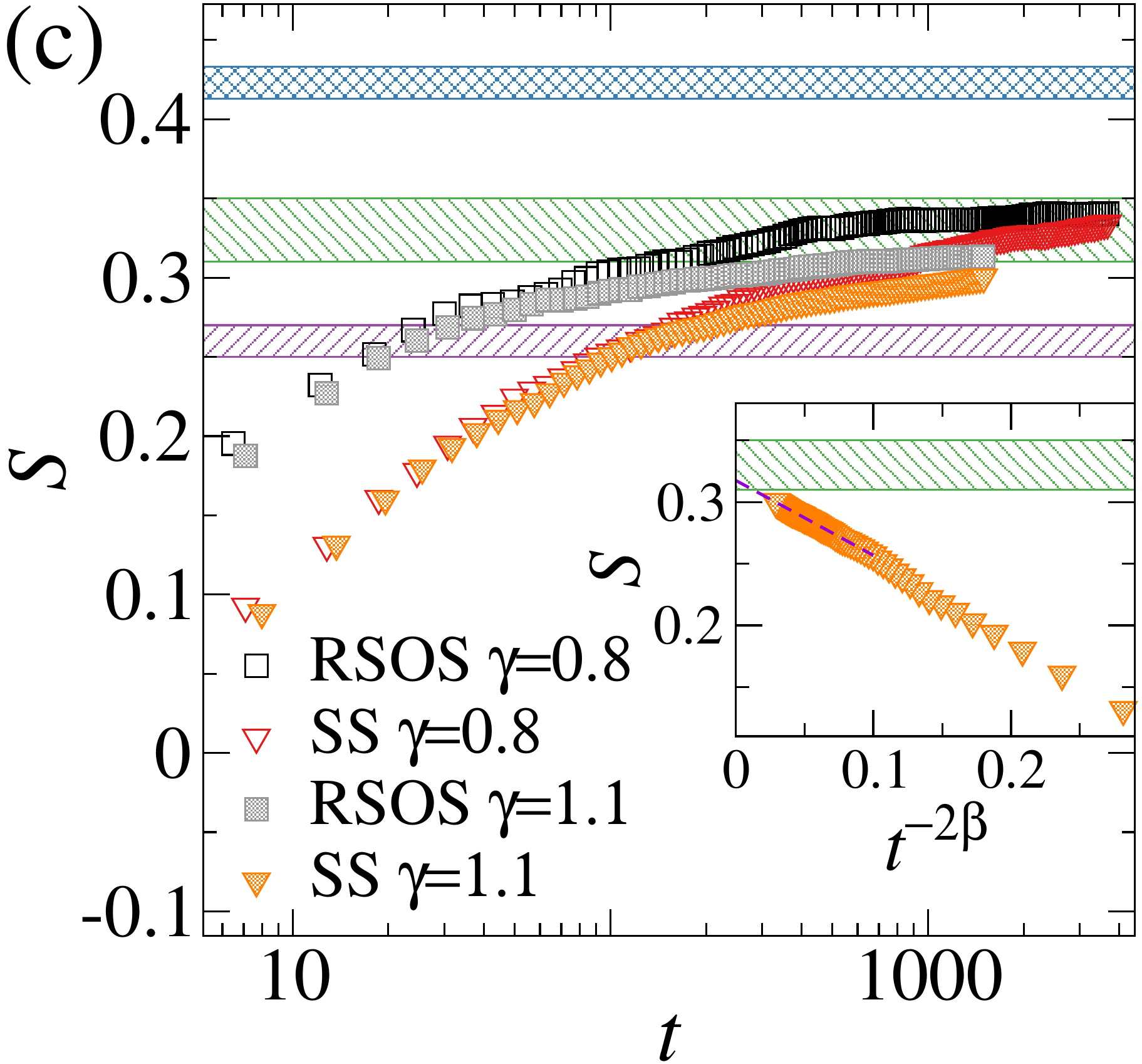} 
	\includegraphics[height=3.9cm]{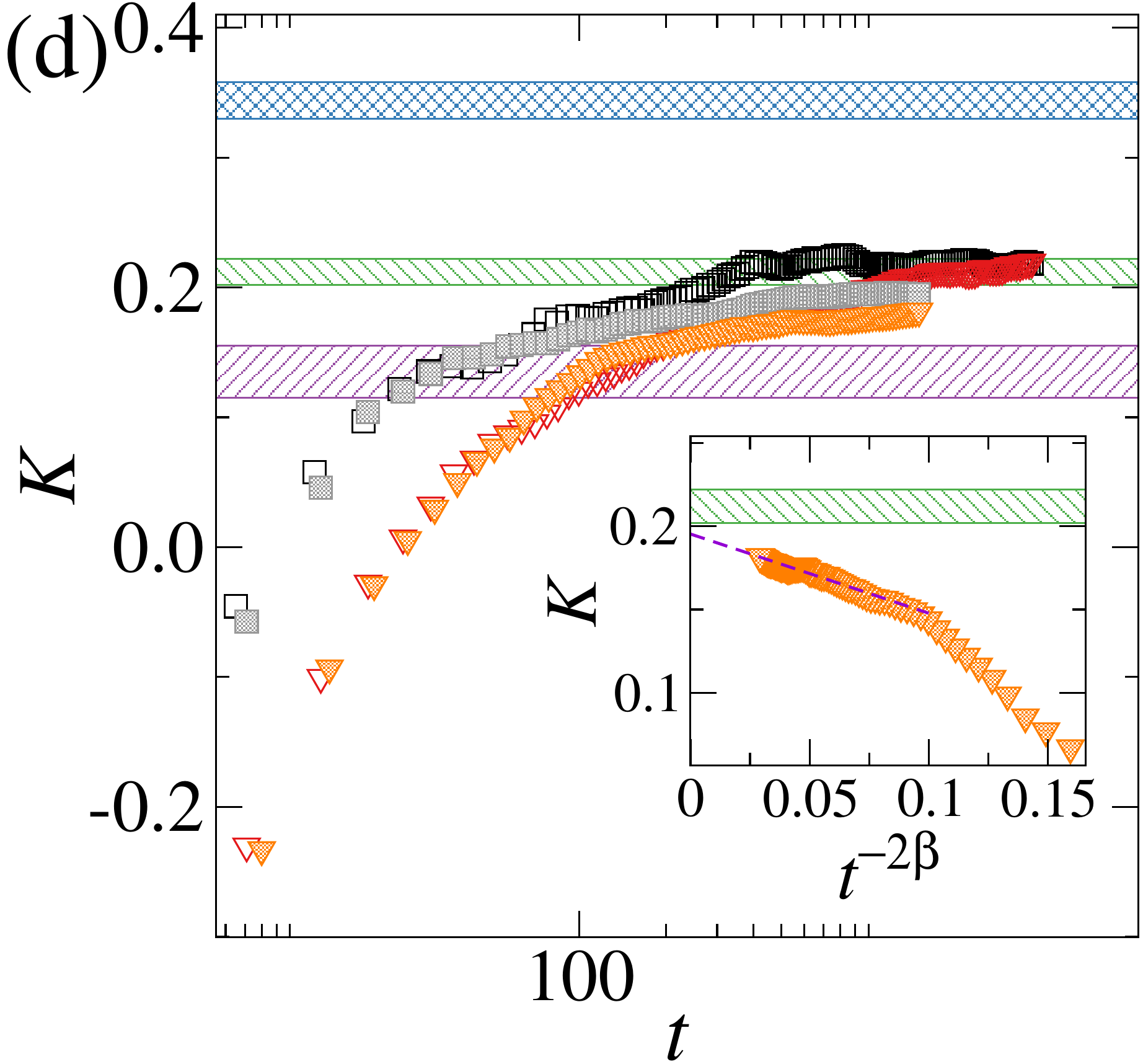} 	
	\caption{Temporal evolution of the skewness $S$ (left) and kurtosis $K$ (right panels) of the HDs for the RSOS and SS models. (a) and (b) show results for $\gamma<1/z$, while in (c) and (d) data for $\gamma>1/z$ are presented, as indicated by the legends. In all panels, the shaded regions represent the ranges of values found in the literature for $S$ and $K$ of the GR HDs in flat and curved geometries, and these moment ratios for the SSR HDs, for the 2D KPZ class. All data shown here are for $\omega=2$, except that for $\gamma=1.1$, for which $\omega=1$.}
	\label{fig3}
\end{figure}

Figures \ref{fig3}(a) and \ref{fig3}(b), respectively, show $S$ and $K$ versus time for systems enlarging slower than the correlation length (i.e., for $\gamma<1/z$). In both graphs, we present the ranges of values numerically established in the literature for the skewness and kurtosis of the GR HDs for flat ($L=const.$) and curved ($L \sim t$) geometries, being $S^{(flat)} = 0.423(9)$, $K^{(flat)}=0.34(1)$, $S^{(curved)} = 0.33(2)$ and $K^{(curved)}=0.21(1)$ \cite{healy12,tiago13,Ismael14}. The ranges for the moment ratios of the HDs for the steady-state regime (SSR) are also shown, where $S^{(ssr)} = 0.26(1)$ and $K^{(ssr)}=0.13(2)$ \cite{Chin,Marinari,Fabio2004kpz,Pagnani}. In Fig. \ref{fig3}(a) we see that at long times the values of $S$ agree with $S^{(ssr)}$, except the result for the SS model with $\gamma=0.2$, which is still a bit smaller than the lower bound for $S^{(ssr)}$ even at the final simulation time. The extrapolation of this data to $t \rightarrow\infty$, however, makes it clear that it converges to the SSR HD skewness [see the insertion in Fig. \ref{fig3}(a)]. A similar behavior is found in the extrapolations (not shown) for all data. In the same way, despite a slower convergence in the data for the SS model, the kurtoses always converge to $K^{(ssr)}$ at long times, as seen in Fig. \ref{fig3}(b). Similar results are found for other values of $\gamma<1/z$ analyzed here, regardless of the $\omega$ considered, strongly suggesting that systems enlarging slower than the correlation length ($\gamma<1/z$) always have the same asymptotic HD as that for the steady-state regime of flat KPZ systems, even though the roughness does not saturate here, as seen in Fig. \ref{fig1}(a).

When $L$ enlarges faster than the correlation length (i.e., with $\gamma>1/z$), we find that both $S$ and $K$ converge to the values expected for the HDs of curved 2D KPZ interfaces (i.e., the 2D counterpart of the TW-GUE distribution), as demonstrated in Figs. \ref{fig3}(c) and \ref{fig3}(d). Note that this is true even for $\gamma>1$. Although the convergence is slower in this case, particularly for the SS model, it is clear from the extrapolations in the insertions that these data also converge to $S^{curved}$ and $K^{curved}$. Similar extrapolations show the same behavior whenever $\gamma>1/z$, for all values of $\omega$ analyzed here. Therefore, this strongly indicates that systems enlarging faster than the correlation length always have the same asymptotic GR HD, which is the one for radial KPZ growth.

\begin{figure}[!t]
	\centering
	\includegraphics[height=3.9cm]{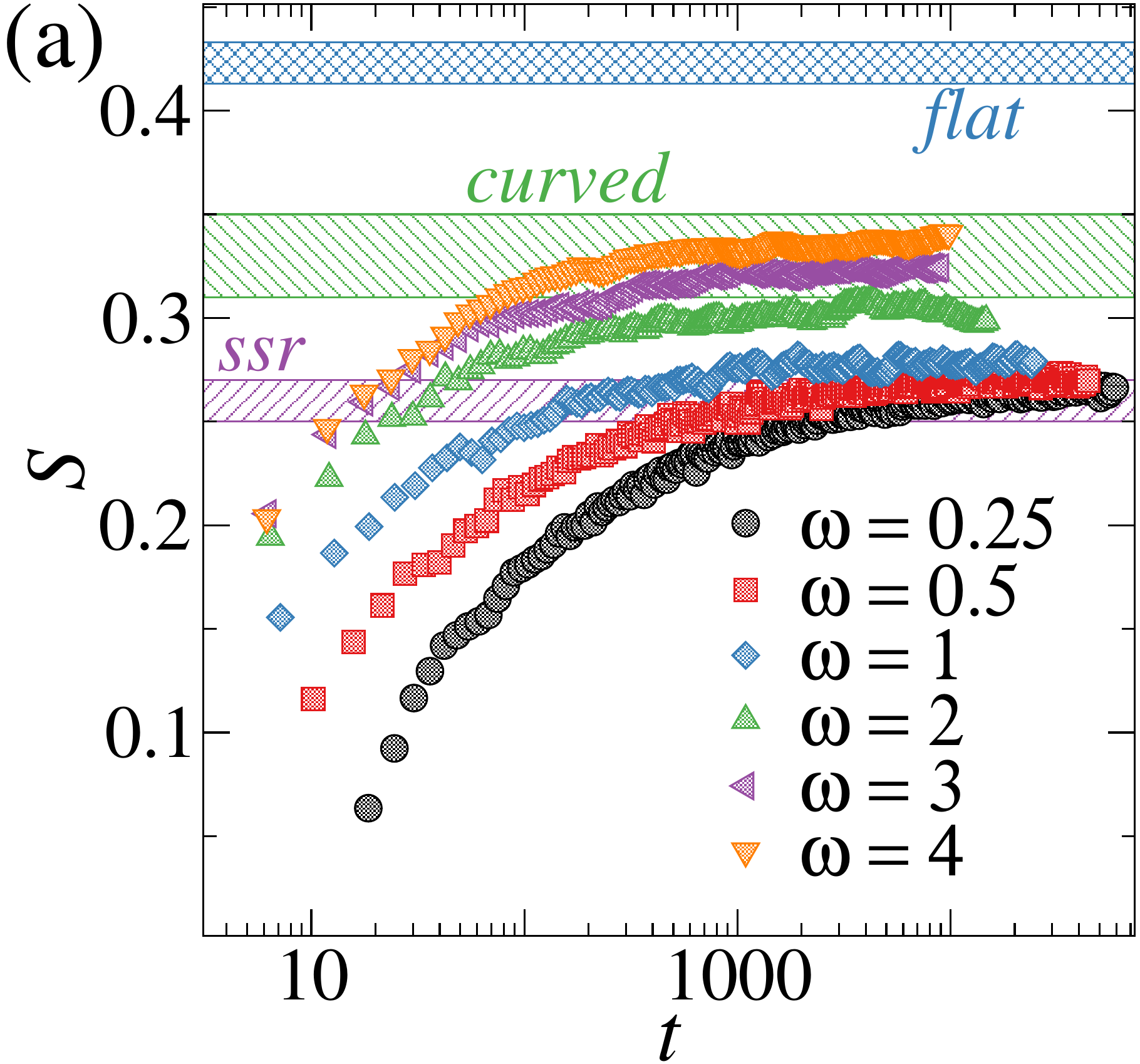} 
	\includegraphics[height=3.9cm]{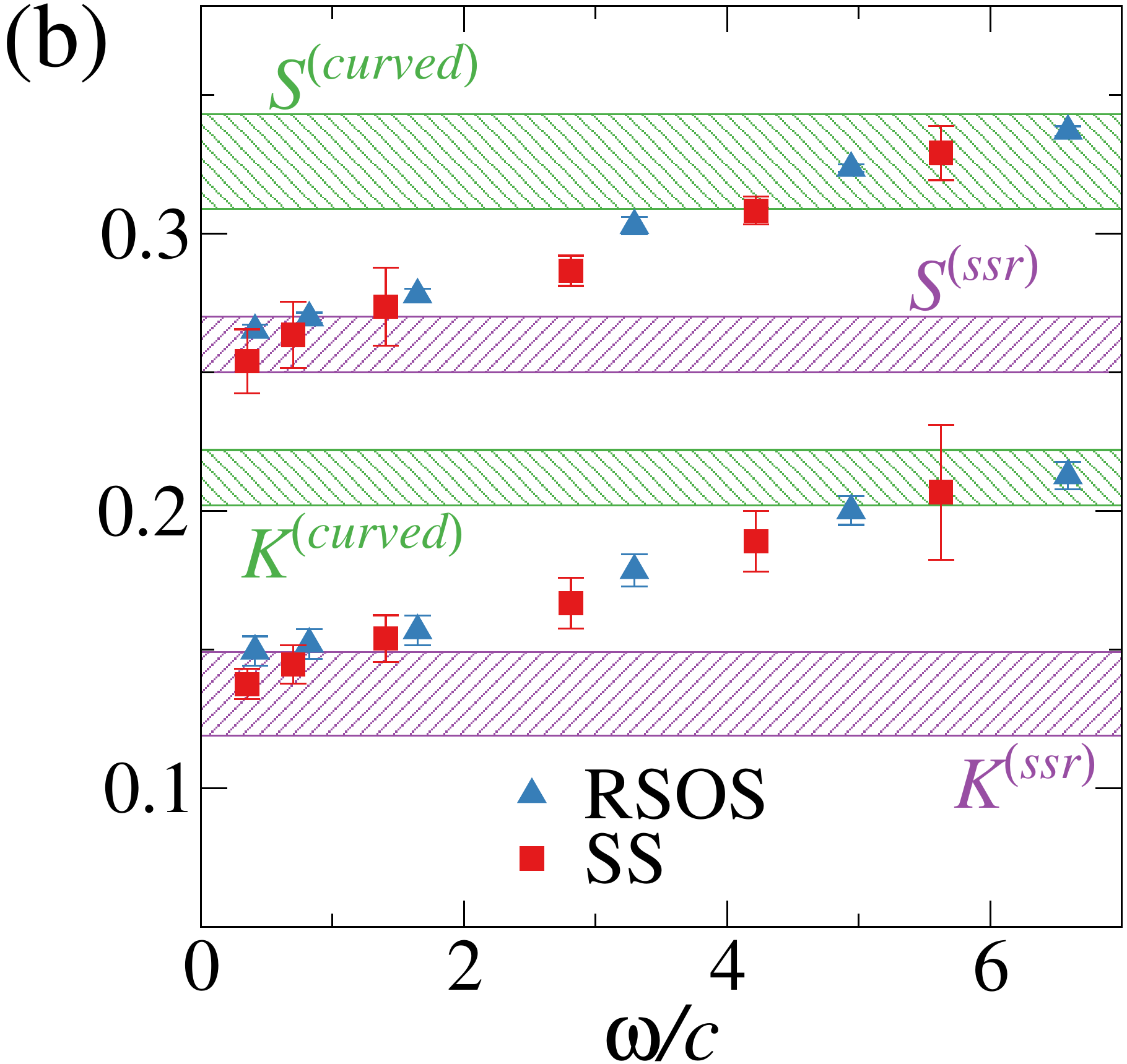} 	
	\caption{(a) Temporal evolution of the skewness $S$ for the RSOS model with $\gamma \approx 1/z$ and different values of $\omega$, as indicated by the legend. (b) Asymptotic values of $S$ (top) and $K$ (bottom) for both models as a function of $\omega/c$. The shaded regions represent the intervals estimated in the literature for $S$ and $K$ for the SSR HD and the GR HDs for flat and curved geometries.}
	\label{fig4}
\end{figure}

Finally, we analyze the case where the substrate enlarges in close competition with the correlation length, namely with $\gamma \approx 1/z \approx 0.620$. When $L(t)=\omega t^{1/z}$ and $\xi(t)=ct^{1/z}$, one may expect that the amplitude $\omega$ shall determine the asymptotic behavior. In fact, Fig. \ref{fig4}(a) shows $S$ versus $t$ for the RSOS model and several values of $\omega$, and one sees that as $\omega$ increases from $0.25$ to $4$, the data converge to a set of values varying from $S^{(ssr)}$ to $S^{(curved)}$. A similar behavior is found for the kurtosis. Since we do not know the exact value of $z$, we have verified that the variation of $\gamma = 0.620$ at the third decimal place (where it is the uncertainty in $1/z$) as a negligible effect on the data when compared with the fluctuations observed in Fig. \ref{fig4}(a). The asymptotic values of $S$ and $K$ [obtained from extrapolations of the data in Fig. \ref{fig4}(a), and analogous ones for $K$, to $t \rightarrow \infty$] for both the RSOS and SS models are depicted in Fig. \ref{fig4}(b) as a function of $\omega/c \approx L(t)/\xi(t)$. This strongly indicates that a universal crossover exists, with a family of asymptotic HDs continuously interpolating between the SSR HD and the one for the 2D curved KPZ subclass, depending solely on the ratio $\omega/c$. We recall that $c=(|\lambda| \sqrt{A})^{1/z}$ for KPZ systems, where $A$ is the amplitude of the height-difference correlation function $G_2(r,t)=\langle \overline{ [h(\vec{x}+\vec{r},t) - h(\vec{x},t)]^2} \rangle \simeq A |\vec{r}|^{2 \alpha}$ \cite{krug1992}. Thereby, from the estimates for $\lambda$ and $A$ reported in Ref. \cite{Ismael14}, we obtained $c=0.61(1)$ and $c=0.71(1)$ for the RSOS and SS models, respectively.

\subsection{Spatial covariances}
\label{secResCov}

\begin{figure}[!t]
	\centering
	\includegraphics[height=4cm]{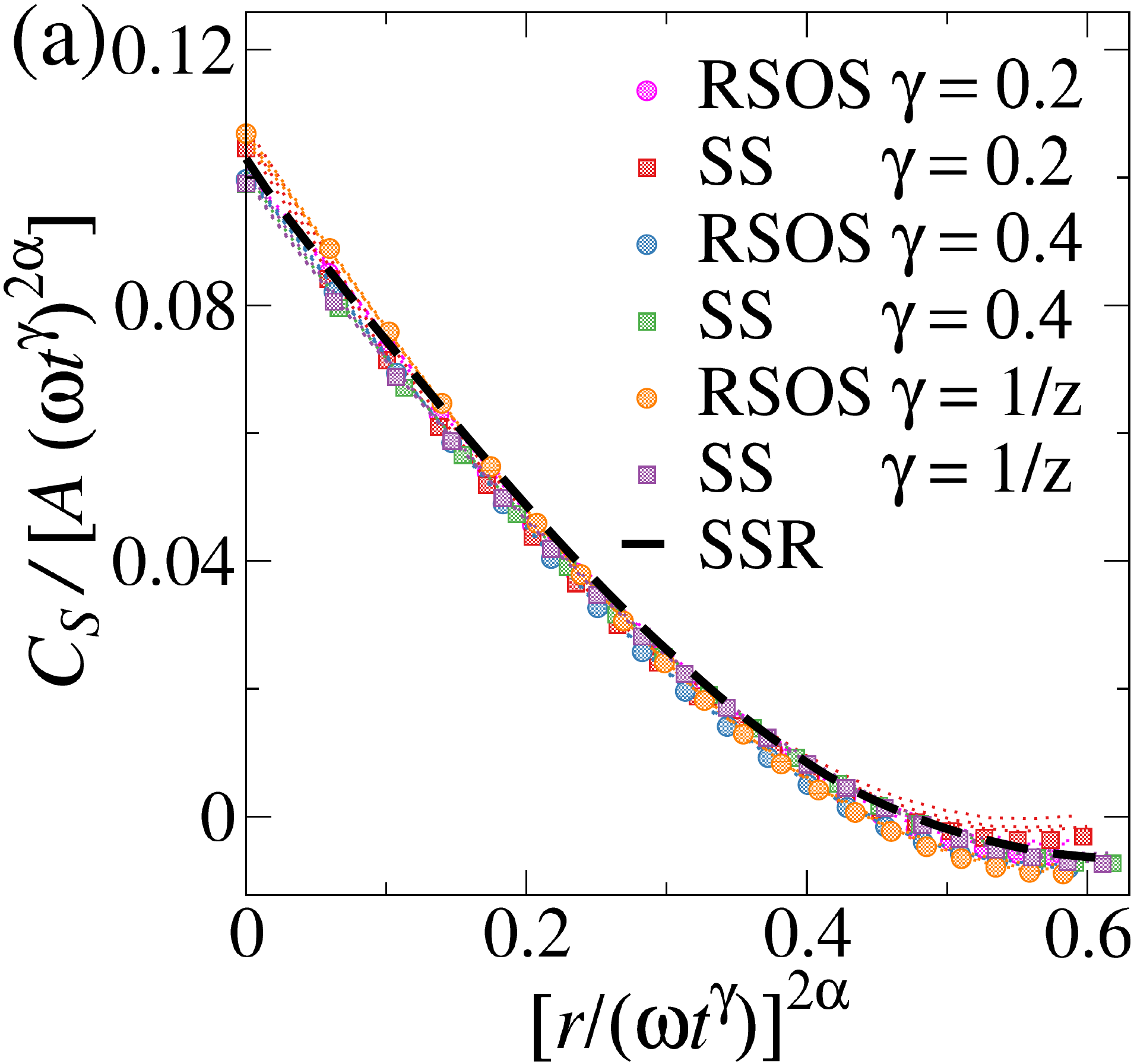} 
	\includegraphics[height=4cm]{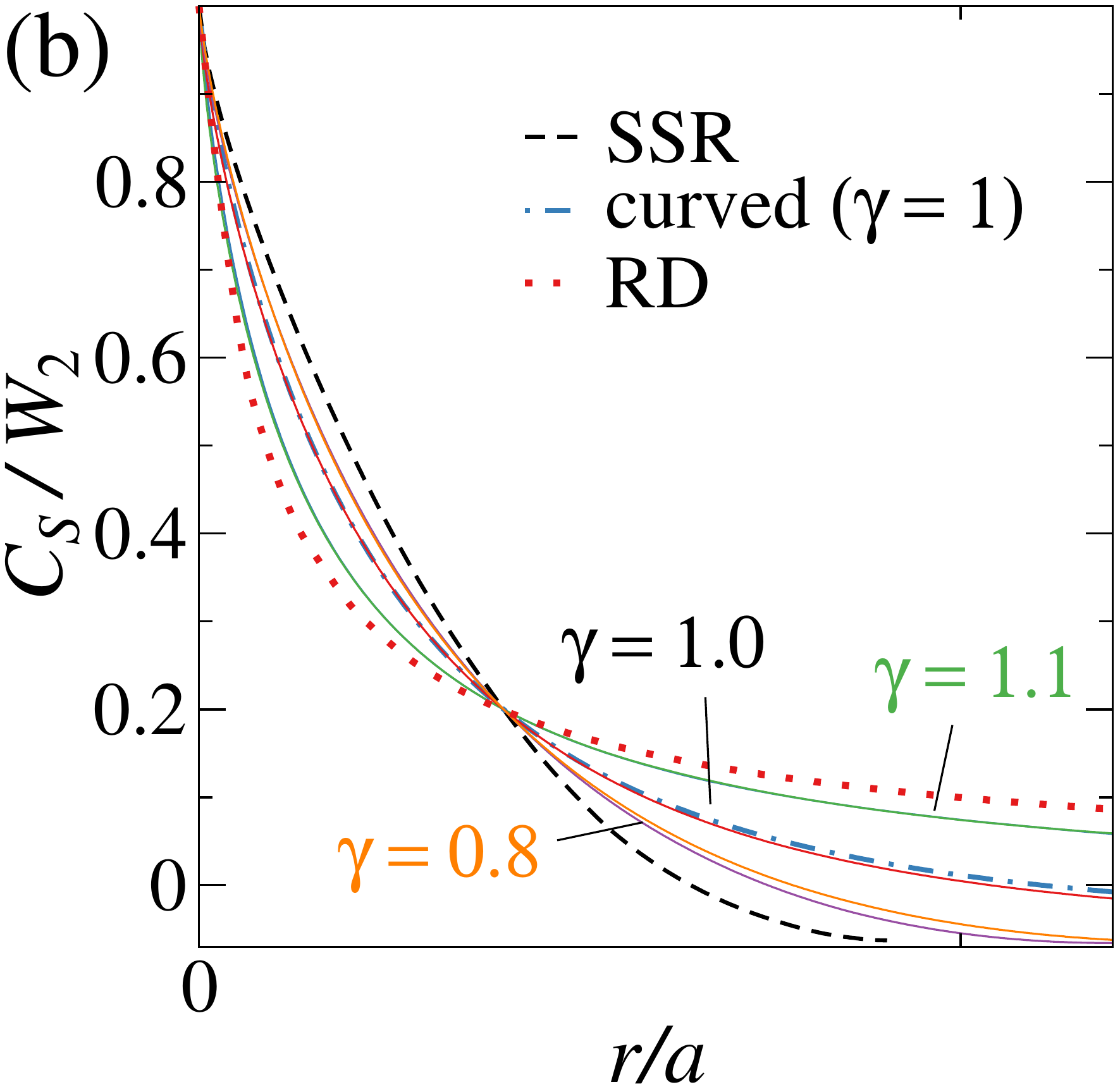}	
	\includegraphics[height=4cm]{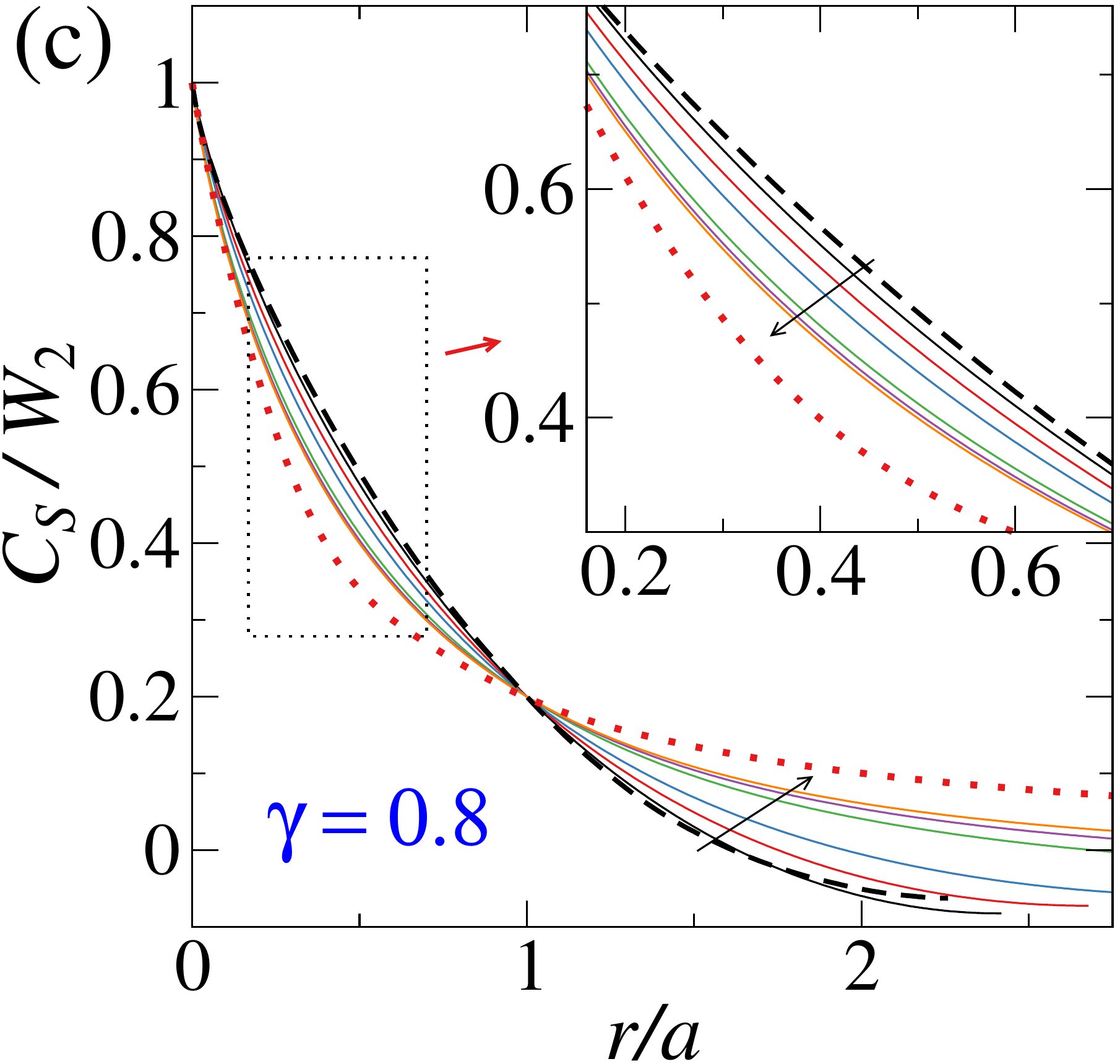}
	\includegraphics[height=4cm]{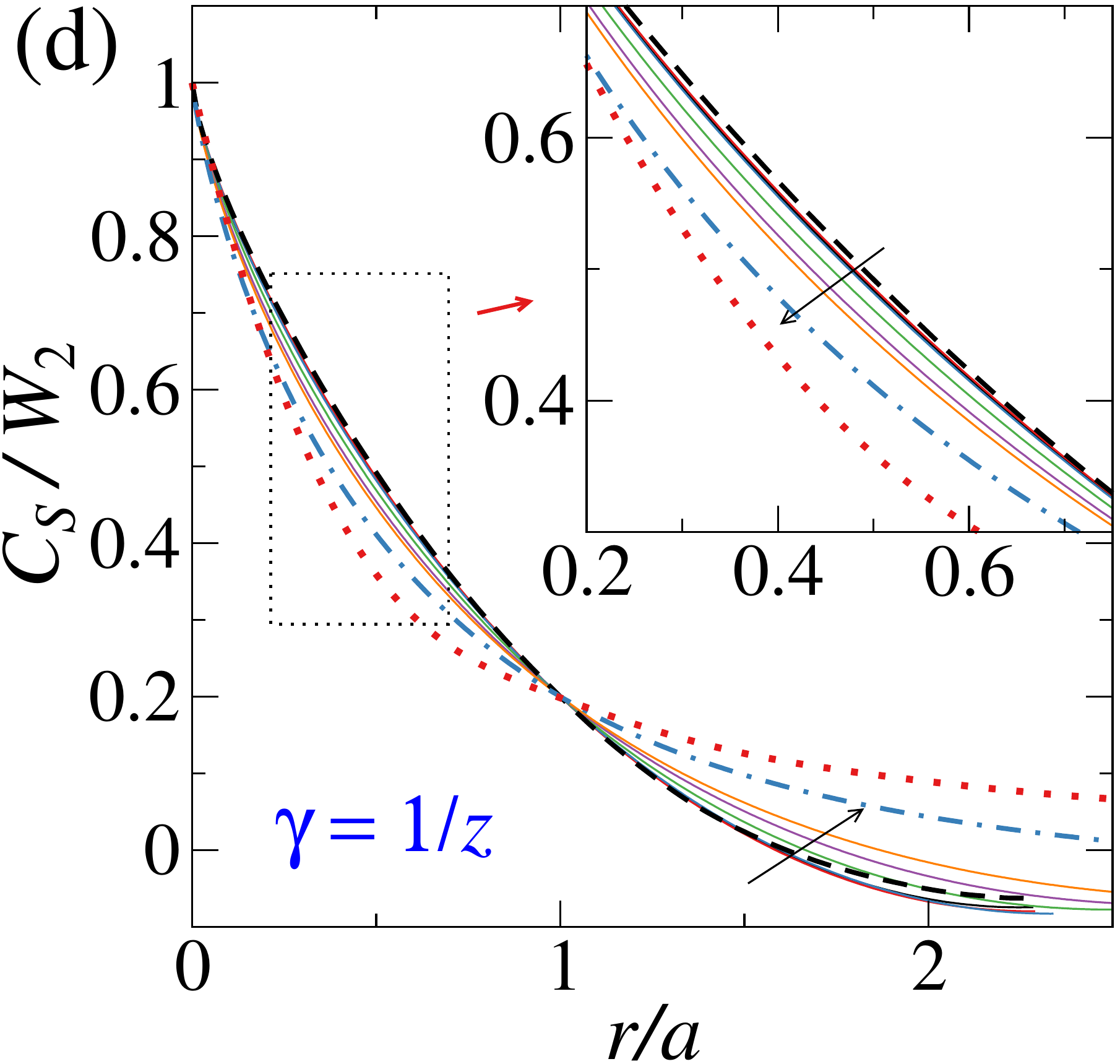}
	\caption{(a) Rescaled spatial covariances $C_S/[A (\omega t^\gamma)^{2\alpha}]$ against $[r/(\omega t^\gamma)]^{2\alpha}$ for both models and several values of $\gamma \le 1/z$, as indicated by the legend. Curves measured at several times are shown, the longest time ones being denoted by symbols. The dashed line is the covariance for the SSR of fixed-size systems, for which $(\omega t^\gamma) \rightarrow L$ was used in rescaling. Rescaled spatial covariances $C_S/W_2$ versus $r/a$ for: (b) both models with $\omega=2$ and the indicated $\gamma$'s; (c) the RSOS model with $\gamma=0.8$ and several $\omega$'s; and (d) the SS model with $\gamma \approx 1/z$ and several values of $\omega$. Results for $\omega \in [0.25 , 4]$, increasing in the direction of the arrows, are shown in (c) and (d), where the insertions highlight the regions inside the boxes in the main panels. Data in (b)-(d) are for times yielding the deposition of $\sim3\times10^9$ particles, which are compared with covariances for the SSR, for the GR in curved geometry ($\gamma=1$), and for a random deposition (RD) on expanding substrates (with $\gamma=1.1$, $\gamma=0.8$ and $\gamma \approx 1/z$, respectively), as indicated by the legend in (b).}
	\label{fig5}
\end{figure}

In this subsection we analyze the effect of the substrate expansion on the spatial covariance, defined in Sec. \ref{secIntro}. In general, it is expected to scale as $C_S(r,t) \simeq W_2 \Psi[x]$, with $\Psi[x]$ assuming different, but universal, forms in the GR and SSR, where $x = r /\xi(t) \sim A r^{2\alpha}/W_2$. Then, since $W_2 \sim A L^{2\alpha} \sim A (\omega t^\gamma)^{2\alpha}$ in the correlated regime found in our systems, we might expect that
\begin{equation}
	C_S(r,t) \simeq A (\omega t^\gamma)^{2\alpha}\Psi_{cr}[(r/\omega t^\gamma)^{2\alpha}],
	\label{eqCsSSR}
\end{equation}
where $\Psi_{cr}[x]$ is a universal scaling function. This is indeed confirmed in Fig. \ref{fig5}(a), which shows rescaled curves of $C_S/[A (\omega t^\gamma)^{2\alpha}]$ versus $(r/\omega t^\gamma)^{2\alpha}$ calculated at the correlated regime, where one observes a striking collapse of data for both models, for different parameters $\omega$ and $\gamma < 1/z$, and for several times. Along with the data for the expanding systems, there is also the curve of $C_S/(AL^{2\alpha})$ versus $(r/L)^{2\alpha}$ calculated at the SSR of the SS model deposited on a square lattice substrate of \textit{fixed} size $L=256$. The agreement of this curve with the other ones demonstrates that the covariance in the correlated regime of the enlarging systems is the same as that for the SSR of the 2D KPZ class.

As an aside, we notice that in the SSR the 1-point height is expected to evolve as $h = \bar{h} + s_{\lambda} A^{\frac{1}{2}}L^{\alpha} \zeta + \cdots$, with $\zeta$ being a fluctuating variable given by the underlying SSR HD \cite{tiago22}. Therefore, since $C_S(r = 0) = W_2 \simeq A (\omega t^{\gamma})^{2\alpha} \langle \zeta^2 \rangle_c$, one shall have the rescaled curves of $C_S/[A (\omega t^\gamma)^{2\alpha}]$ starting at the variance $\langle \zeta^2 \rangle_c$ for long times. From such starting points in Fig. \ref{fig5}(a), we obtain $\langle \zeta^2 \rangle_c =0.103(3)$, which agrees quite well with the value recently found for the SSR HDs of 2D KPZ models deposited on fixed-size substrates: $\langle \zeta^2 \rangle_c =0.1027(5)$ \cite{tiago22}. This provides additional, and very strong, evidence that the height fluctuations (about the mean) for the correlated regime found here and the SSR of flat systems are given by the same HD.

Since the roughness amplitude presents a variation with both $\omega$ and $\gamma$ when $\gamma > 1/z$, to compare the covariances in this case we will adopt the same strategy used in Ref. \cite{Ismael19}, consisting in analyzing $\Psi[y] = C_S/W_2$, with $y=r/a$, where the factor $a$ is chosen to make $C_S/W_2=0.2$ at $r=a$. In this way, all curves start at $\Psi[0]=1$ and coincide also at $\Psi[1]=0.2$, collapsing thus whenever they follow the same universal function. Figure \ref{fig5}(b) presents examples of such rescaled curves, where a clear dependence with $\gamma$ is observed, while data for both models collapse well, particularly for not so large values of $r/a$. Interestingly, these curves depart from close to the SSR one (when $\gamma \rightarrow 1/z$) and approximate to the covariance for a random deposition (RD) performed on substrates expanding as $L \sim t^{\gamma}$ for large $\gamma$. (For long times, the covariances for such RD have a negligible dependence on $t$ and $\omega$, when one considers $L = \omega t^{\gamma}$.) This demonstrates that, when the system enlargement becomes very fast, the correlations generated by the column and row duplications dominate on those coming from particle deposition. When $\gamma$ is not so large, the competition between these two sources of correlations is certainly the reason for the $\gamma$-dependency in $\Psi[y]$. In the same token, we might expect some dependence in $\Psi[y]$ also on $\omega$, for systems expanding faster than $\xi(t)$. This is confirmed in Fig. \ref{fig5}(c), where rescaled covariances for several $\omega$'s are shown, for $\gamma=0.8$, and they are indeed different. Similar results are found for other $\gamma > 1/z$. Note that, just as in Fig. \ref{fig5}(b), the curves start close to the SSR one and move towards the RD covariance (for $\gamma=0.8$) as $\omega$ increases.%, going beyond the $C_S$ of curved interfaces ($\gamma=1$).

Covariances for $\gamma \approx 1/z$ and a small $\omega$ ($=0.25$) are displayed in Fig. \ref{fig5}(a), collapsing quite well with those for the correlated regime ($\gamma< 1/z$). As $\omega$ increases, however, they present a variation, as demonstrated in Fig. \ref{fig5}(d), and seem to tend to the covariance of curved interfaces ($\gamma=1$), once the one for the RD (with $\gamma \approx 1/z$) is considerably far from them. Note that this behavior is similar to what we have seen in Fig. \ref{fig4}(b), where the statistics also changes from the SSR to the GR curved subclass as $\omega$ increases. We remark that only results for the longest deposition times simulated here are shown in Figs. \ref{fig5}(b)-(d), but we have verified that they display only mild finite-time effects, so that the $\omega$- and $\gamma$-dependences observed there are asymptotic features of these spatial covariances.

\section{Conclusion}
\label{secConc}

We have numerically investigated discrete KPZ models deposited on square (on average) substrates, whose lateral size increases as $L = \langle L_x \rangle = \langle L_y \rangle = L_0 + \omega t^{\gamma}$. By changing these parameters, we find a very rich scenario for the asymptotic fluctuations of these expanding systems depending on whether $L$ enlarges faster, at the same rate or slower than the correlation length parallel to the substrate, $\xi \simeq c t^{1/z}$.

For instance, when $\gamma < 1/z$ the surfaces become completely correlated at long times, once $\xi/L \sim t^{1/z - \gamma} \rightarrow \infty$ as $t \rightarrow \infty$, but the roughness does not saturate, because $W_2 \sim L^{2\alpha} \sim t^{2\alpha \gamma}$. Despite this non-stationarity of the roughness, the asymptotic HDs and spatial covariances are consistent with those for the steady-state regime of 2D KPZ surfaces deposited on fixed-size substrates. A careful analysis of the scaling $W_2 \sim t^{2\alpha \gamma}$, correcting it with an intrinsic width and extrapolating effective exponents, allowed us to estimate the roughness exponent of the 2D KPZ class as $\alpha = 0.387(1)$. Considering that $\alpha + z =2$, this gives the dynamic exponent $z=1.613(1)$ and, then, the growth exponent $\beta = \alpha/z = 0.2399(8)$. These values are in remarkable agreement with the best estimates for these exponents in the literature, coming from large-scale simulations performed on fixed-size substrates. For example, our value for $\alpha$ differs by less than 0.03\% from the one estimated by Pagnani \& Parisi \cite{Pagnani} [$\alpha = 0.3869(4)$]. Moreover, our indirect result for $\beta$ is very close to (and agrees within the error bars with) the value obtained from the scaling $W_2 \sim t^{2\beta}$ by Kelling \& \'Odor \cite{Kelling2011} [$\beta=0.2415(15)$]. Notably, in contrast with these works, we are not considering very long times and/or very large substrate sizes --- for some parameters, the maximum sizes, $L_{max}$, attained in our systems are actually small. Notwithstanding, instead of dealing with a few sets of $L$'s, the substrate enlargement naturally samples the roughness for all sizes ranging from $L_0$ to $L_{max}$. This certainly explains the accuracy and robustness (once it is an average over results for different models and several growth conditions) of the $\alpha$ estimated here. One interesting application of this approach is in determining the KPZ exponents for higher dimensions, where simulations are limited to small sizes and short times.

For $\gamma>1/z$ one has $\xi/L \sim t^{1/z - \gamma} \rightarrow 0$ as $t \rightarrow \infty$, so that the growth regime lasts forever, with the squared roughness increasing asymptotically as $W_2 \sim t^{2\beta}$. Importantly, our results strongly indicate that the HDs are asymptotically given by the same distribution previously found in the literature for the 2D curved KPZ subclass, for all $\gamma > 1/z$. The spatial covariances, on the other hand, depend on both $\gamma$ and $\omega$, as a consequence of the correlations introduced by the substrate expansion, which compete with those from the deposition process. This demonstrates that the 1-pt fluctuations are more robust than the 2-pt spatial correlators in these expanding systems. Moreover, the covariance previously found for the 2D curved KPZ subclass seems to be only a particular case (for $\gamma=1$) of a family of continuously varying covariance curves for $L \sim \omega t^{\gamma}$.  

Interestingly, the roughness scaling for the growth and correlated regimes become identical at $\gamma=1/z$, once $W_2 \sim t^{2\alpha\gamma} \sim t^{2\beta}$. In this case, the system stays in a kind of crossover state (between the growth and correlated regime) and a continuous class of HDs can be found depending on the ratio $\omega/c \simeq L/\xi$, which interpolates between the SSR HD (for $\omega/c \rightarrow 0$) and the GR HD of the curved subclass for $\omega/c \gg 1$. The spatial covariances display a similar variation with $\omega/c$, agreeing with that for the SSR when $\omega/c \rightarrow 0$ and moving towards the $C_S$ of curved interfaces as $\omega/c$ becomes large.

The overall scenario above is very similar to the one previously found for KPZ systems deposited on 1D substrates expanding as $L \sim \omega t^{\gamma}$ \cite{Ismael19}. In fact, a diagram summarizing the behaviors found here in terms of $\gamma$ would be very similar to the one reported in Fig. 5 of Ref. \cite{Ismael19}, but with the Gaussian and GUE distributions, as well as the Airy$_2$ covariance replaced by their counterparts for the 2D case. Substantially, this confirms that the behavior of the 1-pt fluctuations and 2-pt correlators of KPZ systems is far richer and interesting than a simple division among few subclasses, also in higher dimensions.

\acknowledgments

The authors acknowledge financial support from CNPq, FAPEMIG and FAPERJ (Brazilian agencies).

\bibliography{bibExpKPZ2D}

\end{document}